\documentclass[reprint,aps,prb]{revtex4-2}

\usepackage[normalem]{ulem}
\usepackage{chemformula} 
\usepackage[T1]{fontenc} 
\usepackage{lmodern}
\usepackage{braket}
\usepackage{amsmath}	
\usepackage{amssymb}
\usepackage{appendix}
\usepackage{hyperref}
\usepackage{epstopdf}
\usepackage{bm}
\usepackage{dcolumn}
\usepackage[version=4]{mhchem}

\usepackage{graphicx}
\usepackage[percent]{overpic} 
\usepackage{color}
\usepackage{transparent} 
\usepackage{cancel}

\usepackage[caption=false]{subfig}
\begin{document} 
	
	\title{Moving skyrmions in Antiferromagnets by Sublattice Displacements}
	
	\author{Michael Lau$^{1,2}$}
	\author{Wolfgang H{\"a}usler$^3$}
	\author{Michael Thorwart$^{1,2}$}
	\affiliation{$^1$I. Institut f{\"u}r Theoretische Physik, University of Hamburg, Notkestra\ss{}e 9, 22607 Hamburg, Germany \\
		$^2$The Hamburg Centre for Ultrafast Imaging, University of Hamburg, Luruper Chaussee 149, 22761 Hamburg, Germany\\
		$^3$ Institut f{\"u}r Physik, Universit{\"a}t Augsburg, Universit{\"a}tsstra\ss{}e 1, 86135 Augsburg, Germany}
	
	\begin{abstract}
	The texture in antiferromagnets hosting a topologically protected skyrmion can be viewed as two effectively coupled ferromagnetic skyrmions. Assuming rigid magnetic configurations, we show that this coupling accounts for the effective mass known for the antiferromagnetic skyrmion and that its dynamics can be viewed as due to a relative displacement of the two sublattice ferromagnetic skyrmions. The theory holds for different antiferromagnetic systems and includes effects from dissipation and external forces caused by electric currents. We verify our analytical results by micromagnetic simulations.
	\end{abstract}
	
	\maketitle

	\section{Introduction}
	\label{sec:Intro}
	
	Magnetic skyrmions are on their way to become technologically relevant elements as magnetic carriers of information \cite{NagaTokOverview}. They are topologically protected quasiparticles in magnetic materials and can be moved by  spin-polarized currents \cite{IwaUniversal} or spin waves \cite{2020Roadmap}. Due to their temperature enhanced random motion skyrmions are  candidates for probabilistic computing devices \cite{TheSkDiff}. So far, ferromagnetic (FM) skyrmions were studied extensively. Currently, there is a trend to study antiferromagnetic (AFM) skyrmions in single- or bilayer forms, the latter having been realized experimentally \cite{Legrand2020}. One advantage of AFM skyrmions is the absence of the skyrmion-Hall effect \cite{BarTre2016}, an effect which makes FM skyrmion motion hard to control. Further, in AFM magnetic stray fields are absent. Also, their enhanced diffusion permits enhanced  thermal activation of the stochastic skyrmion motion \cite{EnhancedDiff}. Moreover, they are found to be driven to comparable velocity by one order of magnitude smaller electric currents than in ferromagnetic systems \cite{Dohi2019}. Hence, AFM skyrmions are more attractive to probabilistic computing, even though they are less thermally stable \cite{AFMStab}. While FM skyrmions only move in the presence of external forces  \cite{KomPap,StierPRL17}, AFM skyrmions exhibit additionally inertial motion due to their effective mass \cite{BarTre2016,Tveten2013staggered,DispPanigrahy}. Its generation is intimately connected to the description of an anti-ferromagnet as two coupled ferromagnets in a mean-field picture, but its precise mechanism demands for further clarification.

	In this work, we formulate a continuum field theory and derive the resulting skyrmion equation of motion from the fundamental Landau-Lifshitz-Gilbert equation (LLG). This explicates how the interaction of the two sublattice ferromagnets generates an effective mass of the AFM skyrmion. We use the Ansatz of sublattice displacements between the FM sublattice skyrmions to explicitly derive the resulting driving torques. A similar technique has been applied previously to skyrmions in the special case of bilayer systems in Ref.\ \cite{DispPanigrahy} by adapting Thiele's equation and phenomenologically considering external forces acting on the skyrmions in each layer. Our work yields an extension beyond bilayer AFMs towards general symmetry classes of antiferromagnets. Instead of extending Thiele's equation and phenomenologically considering external forces, we start from the more fundamental Landau-Lifshitz-Gilbert equation and derive the Thiele equation also containing the aforementioned force without using further assumptions. This way allows us to derive the force explicitly and -- independently -- the energy. We find a natural connection between the force and the kinetic energy, yielding a consistent definition of the skyrmion mass (which, in fact, differs by a factor of 2 compared to Ref.\ \cite{DispPanigrahy}) as the missing link to the kinetic energy.

The interpretation of AFM skyrmion motion in terms of FM sublattices allows us to explain the nontrivial impact of the skyrmion Hall effect (SHE) on the skyrmion motion. In general, the formalism predicts an additional, strong force due to the sublattice skyrmion coupling, providing insights to why AFM skyrmions behave so differently compared to their FM relatives. Some of those scenarios are, for instance, spin wave driven skyrmions \cite{daniels_sk,SkSw}, or skyrmions on curved surfaces \cite{Krav2022}.  Eventually, we consider a spin-polarized electric current applied to each sublattice and derive the current induced motion of the AFM skyrmion for different scenarios. Comparison to numerical simulations confirms our predictions.

	\section{Model}
	\label{sec:Model}
	We consider the continuum model of a two-dimensional antiferromagnet in the $x$-$y$ plane, consisting of two three-dimensional ferromagnetic vector fields  $\bm{a}(\bm{r})$ and $\bm{b}(\bm{r})$ (with $\bm{r}=(x,y)$) of the magnetic moments of the sublattices A and B, respectively. The energy functional \cite{BRWM2002}
	\begin{equation}
		\label{eq:conti_energy_afm}
		\begin{aligned}
			W &= \int \text{d}^2 \bm{r} \big\{ \frac{\lambda}{2} \bm{a}\bm{b} + \frac{A^{\prime}}{2} \sum_{\mu \in \{x,y\}} \left[ \left( \partial_\mu \bm{a} \right)^2 + \left( \partial_\mu \bm{b} \right)^2 \right]\\
			&+ A \sum_{\mu \in \{x,y\}} \left[\left(\partial_\mu \bm{a} \right) \left(\partial_{\mu} \bm{b} \right) \right] + D \bm{a} \cdot (\bm{\nabla} \times \bm{b})\\
			&+ \frac{D^{\prime}}{2} \left[ \bm{a} \cdot (\bm{\nabla} \times \bm{a}) + \bm{b} \cdot (\bm{\nabla} \times \bm{b}) \right] - \frac{K^{\prime}}{2}(a_z^2 + b_z^2)\\
			&- K a_z b_z \big\} \, 
		\end{aligned}
	\end{equation}
	includes the homogeneous ($\lambda$) and inhomogeneous ($A$, $A^{\prime}$) parts of the exchange interaction, the Dzyaloshinskii-Moriya interactions ($D^{\prime}$ and $D$), as well as the magnetocrystalline anisotropy ($K^{\prime}$ and $K$) in $z$-direction, see \cite{BRWM2002} for more details. We assume that the magnetic moments $\bm{n} \in \{\bm{a}, \bm{b}\}$ evolve according to the Landau-Lifshitz-Gilbert equation (LLG) \cite{LL, Gilbert} $\partial_t \bm{n} = -\gamma \bm{n} \times \bm{H}_{\text{eff}} + \alpha \bm{n} \times \partial_t \bm{n}$ on each sublattice, so that we obtain a set of two coupled equations of motion (assuming a common Gilbert damping parameter $\alpha$ \cite{GomoCurAFM}, as well as setting the gyromagnetic ratio to $\gamma = 1$). Their coupling arises due to the effective fields $\bm{H}_{\text{eff}}^{A} = -\delta W / \delta \bm{a}$ and $\bm{H}_{\text{eff}}^{B} = -\delta W / \delta \bm{b}$, where for sublattice A
	\begin{equation}
		\label{eq:Heff_conti}
		\begin{aligned}
			\bm{H}_{\text{eff}}^{A} = &-\frac{\lambda}{2} \bm{b} + A \bm{\nabla}^2 \bm{b} - D (\bm{\nabla} \times \bm{b}) + K b_z \bm{z}\\
			&+A^{\prime} \bm{\nabla}^2 \bm{a} - D^{\prime} (\bm{\nabla} \times \bm{a}) + K^{\prime} a_z \bm{z},
		\end{aligned}
	\end{equation}
	while $\bm{H}_{\text{eff}}^{B}$ is obtained from Eq.\ (\ref{eq:Heff_conti}) after interchanging all $\bm{a}$ and $\bm{b}$ vectors. Each separates into intra- and inter-sublattice fields. The intra-sublattice effective field,
	\begin{equation}
		\label{eq:intra_heff_A}
		\bm{H}_{\text{intra}}^A = A^{\prime} \bm{\nabla}^2 \bm{a} - D^{\prime} (\bm{\nabla} \times \bm{a}) + K^{\prime} a_z \bm{z},
	\end{equation}
	only consists of terms depending on the same sublattice while the inter-sublattice effective field,
	\begin{equation}
		\label{eq:inter_heff_A}
		\bm{H}_{\text{inter}}^A = -\frac{\lambda}{2} \bm{b} + A \bm{\nabla}^2 \bm{b} - D (\bm{\nabla} \times \bm{b}) + K b_z \bm{z},
	\end{equation}
	consists of term only depending on the respective other sublattice. One AFM skyrmion can therefore be considered as being composed from two FM skyrmions, one residing on each sublattice.
	
	\section{Sublattice skyrmion displacement} 
	\label{sec:displacement}
	When the two sublattice skyrmions, centered at $\bm{R}_A$ and $\bm{R}_B$, respectively, are displaced by $\bm{\delta} = \bm{R}_A - \bm{R}_B$ relative to each other, a driving force results on the AFM skyrmion, as we show in the following. Without loss of generality, we choose the FM skyrmion on sublattice A as a reference. We then consider a  magnetic structure $\bm{b}$ weakly displaced spatially by some given small displacement vector $\bm{\delta}=(\delta_x, \delta_y)$, relative to its original position in the AFM skyrmion. This displacement may be considered as virtual, but may as well have some real physical presence in some cases (see below). Thus, we  may approximate
	\begin{equation}
		\label{eq:taylor_sl_B}
		\bm{b}(\bm{r}) \approx -\bm{a}(\bm{r}) - (\bm{\delta} \bm{\nabla}) \bm{a}(\bm{r}) - \frac{1}{2} \sum_{\mu, \nu=x,y} \delta_{\mu} \delta_{\nu} (\partial_\mu \partial_\nu \bm{a}(\bm{r}))  \, .
	\end{equation}
	A similar approximation was used for studying skyrmion motion in bilayer systems \cite{DispPanigrahy}. Here we use it to study an antiferromagnetic system consisting of two sublattice vector fields, as in Ref.\ \cite{BRWM2002}. When taking sublattice B as reference, the same displacement can be expressed by approximating the sublattice A field as
	 \begin{equation}
	 	\label{eq:taylor_sl_A}
	 	\bm{a}(\bm{r}) \approx -\bm{b}(\bm{r}) + (\bm{\delta} \bm{\nabla}) \bm{b}(\bm{r}) - \frac{1}{2} \sum_{\mu, \nu=x,y} \delta_{\mu} \delta_{\nu} (\partial_\mu \partial_\nu \bm{b}(\bm{r}))  \, .
 \end{equation}
	We note that the sign of $\bm{\delta}$ changes whether sublattice A or B is taken as reference lattice. In real systems, such a displacement can arise due to various reasons. Examples are AFM skyrmions driven by external currents or by antiferromagnetic spin waves, which act differently on each sublattice \cite{SkSw} causing the magnetic moments of each sublattice to oscillate with a different amplitude \cite{KefKit,SkSw}. Therefore, each sublattice skyrmion experiences forces of different strength which spatially displaces them.
	
	Although skyrmions may exhibit a variety of internal excitation modes \cite{AFMskyrmionModes}, we restrict ourselves to rigid sublattice skyrmions and solely use their centers of mass $\bm{R}_{\text{A/B}}(t)$ (with $\bm{R}_{\text{A}}(t)-\bm{R}_{\text{B}}(t)=\bm{\delta}(t)$) as collective coordinates \cite{CollectiveCoordinates}. Thiele's equation \cite{Thiele} can be generalized to drift velocities $\bm{v}^{A/B}(t)$ of arbitrary time dependencies, yielding $\partial_t \bm{a} = -\bm{v}^A(t) \bm{\nabla} \bm{a}$ and $\partial_t \bm{b} = - \bm{v}^B(t) \bm{\nabla} \bm{b}$. Applying this to the LLG and integrating over space leads to
	\begin{equation}
		\label{eq:sl_thiele}
		\begin{aligned}
			4 \pi Q \sum_{\nu=x,y} v_{\nu}^A \epsilon_{\mu \nu z} &- \alpha \sum_{\nu=x,y} v_{\nu}^A \int \left(\partial_{\mu} \bm{a}\right) \cdot \left(\partial_\nu \bm{a}\right) \text{d}^2 \bm{r}\\ 
			&-  \int \left(\partial_\mu \bm{a} \right) \cdot \bm{H}_{\text{eff}}^{A}\ \text{d}^2 \bm{r}= 0,
		\end{aligned}
	\end{equation} 
	with the Levi-Civita symbol $\epsilon_{\mu \nu \kappa}$. This is the Thiele equation $-\bm{G} \times \bm{v} - \alpha \bm{\mathcal{D}} \bm{v} - \nabla V(\bm{r})=0$ for a FM skyrmion \cite{IwaConstricted} with the dissipation tensor $\bm{\mathcal{D}}$ and the gyrocoupling $\bm{G} = 4 \pi Q \bm{z}$ \cite{Thiele,EverGarPRB2012}, where  $Q$ denotes the topological quantum number of a skyrmion. We consider here only skyrmions with a quantum number $Q = \pm 1$. The potential $V(\bm{r})$ is created by the displaced skyrmion in sublattice B and leads to a force
	\begin{equation}
		\label{eq:force_AB}
		\bm{F}_{AB} = \int \left(\partial_\mu \bm{a} \right) \cdot \bm{H}_{\text{eff}}^{A}\ \text{d}^2 \bm{r}
	\end{equation}
	acting on sublattice A. The same can be done for the other sublattice B, yielding the force
	\begin{equation}
		\label{eq:force_BA}
		\bm{F}_{BA} = \int \left(\partial_\mu \bm{b} \right) \cdot \bm{H}_{\text{eff}}^{B}\ \text{d}^2 \bm{r}
	\end{equation}
	acting on the skyrmion in sublattice B. The effective sublattice field $\bm{H}_{\text{eff}}^A$ separates into terms solely depending on the other sublattice B, e.g., $A \bm{\nabla}^2\bm{b}$, which we call {\em inter}-sublattice terms $\bm{H}_{\text{inter}}^A$ , and terms solely depending on the same sublattice A, e.g., $K^{\prime} a_z \bm{z}$, which we call {\em intra}-sublattice terms	$\bm{H}_{\text{intra}}^A$, cf.\ Eqs.\ (\ref{eq:inter_heff_A}) and (\ref{eq:intra_heff_A}). Ignoring distortions of the FM skyrmion occuring with displacements, in the spirit of Thiele's approximation, intra-sublattice terms are independent of a sublattice displacement. For the analytical calculations, we use spherical coordinates, $\bm{n}=(\sin\theta\:\cos\Phi,\sin\theta\:\sin\Phi,\cos\theta)$, and assume the typical form of the sublattice skyrmion profile \cite{NagaTokOverview}, i.e., a radially symmetric profile with $\theta = \theta(\rho)$ as well as $\Phi = m \varphi + \varphi_0$. Here, and $m=\pm 1$ distinguishes skyrmions from Anti-skyrmions, while the helicity $\varphi_0=0$ (Néel) or $\varphi_0=\pi/2$ (Bloch) controls the Skyrmion type. Explicit integration over the azimuthal coordinate $\varphi$ then leads to
	\begin{equation}
		\label{eq:intra_integrals}
		\begin{aligned}
			&\int_{0}^{2 \pi} \left( \partial_{\mu} \bm{n} \right) \cdot \left( \bm{\nabla}^2 \bm{n} \right) \text{d} \varphi = 0\, , \\
			&\int_{0}^{2 \pi} \left( \partial_{\mu} \bm{n} \right) \cdot \left( \bm{\nabla} \times \bm{n} \right) \text{d} \varphi = 0\, ,\\
			&\int_{0}^{2 \pi} \left( \partial_{\mu} \bm{n} \right) \cdot \left( n_z \bm{z} \right) \text{d} \varphi = 0\, ,
		\end{aligned}
	\end{equation}
	for three intra-sublattice terms, where $\bm{n} \in \{\bm{a}, \bm{b}\}$ is the general representation of a skyrmion, so that Eq.\ \eqref{eq:intra_integrals} holds for both sublattices. Although, we explicitly considered here a bulk DMI and a Bloch-type skyrmion, $\varphi_0 = \pi/2$, the results are similar for interfacial DMI and assuming a Néel-type skyrmion, $\varphi_0 = 0$. Hence, we conclude that our formalism applies to all radially symmetric  magnetic structures in antiferromagnets containing only these types of intra-sublattice terms, i.e., $\int (\partial_\mu \bm{a} ) \cdot \bm{H}_{\text{intra}}^{A}\ \text{d}^2 \bm{r}= 0$. The corresponding statement for sublattice B holds. Thus, only the inter-sublattice terms, see Eq.\ \eqref{eq:inter_heff_A}, contribute to the force $\bm{F}_{AB} = \int \left(\partial_\mu \bm{a} \right) \cdot \bm{H}_{\text{inter}}^{A}\ \text{d}^2 \bm{r} $. These depend on the displacement and can be expressed by inserting Eq.\ (\ref{eq:taylor_sl_A}) with Eq.\ (\ref{eq:Heff_conti}) into Eq.\ \eqref{eq:force_AB}. In consequence only terms from a single sublattice appear so that the integration over this sublattice can be carried out. By the Taylor expansion to quadratic order, the relevant integrals are
		\begin{equation}
		\label{eq:inter_integrals}
		\begin{aligned}
			&\int_{0}^{\infty} \int_{0}^{2 \pi} \left( \partial_{\mu} \bm{b} \right) \cdot \bm{H}_{\text{inter}}^A \text{d} \varphi\ \text{d} \rho = 0\, , \\
			&\int_{0}^{\infty}\int_{0}^{2 \pi} \sum_{\nu} \delta_{\nu} \left( \partial_{\nu} \partial_{\mu} \bm{b} \right) \cdot \bm{H}_{\text{inter}}^A \text{d} \varphi\ \text{d} \rho = \Lambda \delta_{\mu}\, , \\
			&\int_{0}^{\infty}\int_{0}^{2 \pi} \sum_{\nu, \nu'} \delta_{\nu'} \delta_{\nu} \left(\partial_{\nu'} \partial_{\nu} \partial_{\mu} \bm{b} \right) \cdot \bm{H}_{\text{inter}}^A \text{d} \varphi\ \text{d} \rho = 0\, ,
		\end{aligned}
	\end{equation}
	where $\Lambda$ is a constant expressed in units of force per unit length, analogous to the spring constant in Hooke's Law. Consequently, the force term appearing in Eq.\ \eqref{eq:sl_thiele} is
	\begin{equation}
		\label{eq:lambda_def}
		\begin{aligned}
			\int \left(\partial_\mu \bm{a} \right) \cdot \bm{H}_{\text{eff}}^{A}\ \text{d}^2 \bm{r} = \Lambda \delta_\mu \, ,
		\end{aligned}
	\end{equation} 
	proportional to the displacement $\bm{\delta}$, with the positive force constant $\Lambda$. Since the radial profile $\theta(\rho)$ of the skyrmion cannot be expressed analytically, the integration over $\rho$ remains to be done numerically. In Appendix \ref{appendix2} the explicit form of this integral is given. Similar steps for $\bm{b}$ yield $\int \left(\partial_\mu \bm{b} \right) \cdot \bm{H}_{\text{eff}}^{B}\ \text{d}^2 \bm{r} = -\int \left(\partial_\mu \bm{a} \right) \cdot \bm{H}_{\text{eff}}^{A}\ \text{d}^2 \bm{r}$, confirming the assumption \cite{DispPanigrahy} of opposite inter-sublattice forces $\bm{F}_{AB} = -\bm{F}_{BA}$. The dissipation tensor $\bm{\mathcal{D}}$ is the same for each sublattice and isotropic in the $x$-$y$-plane, such that it becomes a constant and could be absorbed into the Gilbert damping. Using the fact that the opposite polarization of the two sublattice skyrmions leads to opposite topological charges $Q_A = -Q_B$, where we consider sublattice A as reference with $Q = Q_A$, we can construct two Thiele equations coupled by the force $\bm{F} = -\Lambda \bm{\delta}$
	\begin{equation}
		\label{eq:thiele_sk_sl}
		\begin{aligned}
			-\bm{G} \times \bm{v}_A - \alpha \mathcal{D} \bm{v}_A - \Lambda \bm{\delta} &= 0\, , \\
			\bm{G} \times \bm{v}_B - \alpha \mathcal{D} \bm{v}_B + \Lambda \bm{\delta} &= 0\,.
		\end{aligned}
	\end{equation}
	Then, Thiele's equation can be rearranged, leading to
	\begin{equation}
		\label{eq:Thiele_A_B_damp}
		\bm{v}^{A(B)} = \bm{z} \times \left( \frac{\Lambda}{4 \pi Q} \bm{\delta} \pm \frac{\alpha \mathcal{D}}{4 \pi Q} \bm{v}^{A(B)} \right) \, .
	\end{equation}
	Accordingly, in the absence of damping, $\alpha=0$, the velocity $\bm{v}^A = \bm{v}^B \equiv \bm{v}$ is perpendicular to the sublattice displacement $\bm{\delta}$. Sublattice skyrmions move in parallel and their distance $\bm{\delta}$ remains constant over time which, in turn, yields a constant $\bm{v}$. 
	
	To include finite damping $\alpha >0$ into the calculations, we work with the velocities of the sublattice skyrmions, $\bm{v}_A$ and $\bm{v}_B$, similar as in  a mechanical two-body problem. The total velocity of the AFM skyrmion is, akin to a center-of-mass velocity, $\bm{v} = (\bm{v}_A + \bm{v}_B) /2$. The relative velocity $\Delta \bm{v} = \bm{v}_A - \bm{v}_B$ is related to the time derivative of the displacement, i.e., $\partial_t \bm{\delta} = \bm{v}_A - \bm{v}_B$. In these terms, Eq.\ \eqref{eq:Thiele_A_B_damp} can be rewritten in the form
	\begin{equation}
		\label{eq:two-eq-delta-v}
		\begin{aligned}
		4 \pi Q \bm{z} \times \partial_t \bm{\delta} + 2 \alpha \mathcal{D} \bm{v} &= 0\, ,\\
		\frac{(4 \pi Q)^2}{\alpha \mathcal{D}} \partial_t \bm{\delta} + \alpha \mathcal{D} \partial_t \bm{\delta} + 2 \Lambda \bm{\delta} &= 0\, ,
	\end{aligned}
	\end{equation}
	where the second line is an ordinary differential equation for  $\bm{\delta}(t)$. Assuming an initial displacement $\bm{\delta}_0$, its solution is $\bm{\delta}(t) = \bm{\delta}_0 \exp\left[\frac{-2 \Lambda \alpha \mathcal{D}}{16 \pi^2 + \alpha^2 \mathcal{D}^2} t\right]$. Inserting the solution $\bm{\delta}$ into Eq.\ \eqref{eq:Thiele_A_B_damp} yields the AFM skyrmion velocity
	\begin{equation}
		\label{eq:v_AFM_SK_damp}
		\bm{v}(t) = \bm{z} \times \frac{4 \pi Q \Lambda \delta_0}{16 \pi^2 + \alpha^2 \mathcal{D}^2 }  \exp\left[ \frac{-2 \Lambda \alpha \mathcal{D}}{16 \pi^2 + \alpha^2 \mathcal{D}^2} t \right]\, . 
	\end{equation}
	 
	For small damping, $\alpha\mathcal{D} \ll 4 \pi$, the $x$-component of the skyrmion velocity in Eq.\ \eqref{eq:v_AFM_SK_damp} can be considered as $\dot{x}(t) = v_0 \exp\left[ -\left(\frac{\Gamma}{m}\right) t \right]$, where $v_0$ is determined by $\delta_0$ according to Eq.\ \eqref{eq:Thiele_A_B_damp}. This is precisely the solution of the equation of motion of a damped massive particle $m \ddot{x} + \Gamma \dot{x} = 0$ with an initial velocity $v_0$. In agreement with Ref.\ \cite{DispPanigrahy}, we define the dissipation constant  as $\Gamma = 2 \alpha \mathcal{D}$ and the effective mass as $m = 16 \pi^2 / \Lambda$.
	
	For the special case of an antiferromagnetically coupled bilayer system \cite{KoshBilayer} the on-site form $J_{\text{inter}} \sum_i \bm{a}_i \cdot \bm{b}_i$ is taken as coupling between sublattice sites, which in the continuum we approximate by $(\lambda/2) \bm{a}(\bm{r}) \cdot \bm{b}(\bm{r})$ at same $\bm{r}$. This leads to the force constant $\Lambda = (\lambda/2) \mathcal{D}$ and, therefore, to the effective mass being inversely proportional to the interlayer coupling. Formally, in the limit of vanishing interlayer coupling, an infinite mass arises which cannot be accelerated by finite forces. This limit poses no contradiction, since for vanishing interlayer coupling, both sublattice skyrmions indeed cannot exert forces upon one another so that no motion of the center of mass can be induced.

	\subsection{Energy considerations}
In order to derive the excitation energy induced by the sublattice skyrmion displacement, we consider the energy from Eq.\ \eqref{eq:conti_energy_afm}. A derivation for more general situations can be found in Appendix \ref{apendix1}. Since we assume rigid sublattice skyrmions, only their displacement contributes to the excitation energy. In other words, only the coupling energy $W_c$ consisting of both sublattices contributes while all terms consisting of only one sublattice do not depend on the displacement. This refers to terms like $W_0^A = -\frac{1}{2}\int \bm{a} \cdot \bm{H}_{\text{intra}}^A\ \text{d}^2 \bm{r}$ (and similarly for $W_0^B$) in Eq.\ \eqref{eq:conti_energy_afm} which depend only on one sublattice each, with the intra sublattice effective field taken from Eq.\ \eqref{eq:intra_heff_A}. They do not contribute to the excitation energy. On the other hand, $W_c = -\int \bm{a} \cdot \bm{H}_{\text{inter}}^A\ \text{d}^2 \bm{r}$ contains different sublattices due to inter-sublattice terms from the effective field, Eq.\ \eqref{eq:inter_heff_A} and therefore contributes to the displacement excitation. We approximate the displacement in the vector field $\bm{a}$ as in Eq.\ \eqref{eq:taylor_sl_A}. By this, the integration appears now only over one sublattice, here B, and yields
	\begin{equation}
		\label{eq:coupling_energy_parts}
		\begin{aligned}
			W_c = &\int \bm{b} \cdot \bm{H}_{\text{inter}}^A\ \text{d}^2 \bm{r}
			-\int \sum_{\mu} \delta_{\mu} \left(\partial_{\mu} \bm{b}\right)\cdot \bm{H}_{\text{inter}}^A\ \text{d}^2 \bm{r}\\
			&+\int \frac{1}{2} \sum_{\mu, \nu} \delta_{\mu} \delta_{\nu} \left(\partial_{\mu} \partial_{\nu} \bm{b}\right)\cdot \bm{H}_{\text{inter}}^A\ \text{d}^2 \bm{r}\;.
		\end{aligned}
	\end{equation}
The first term describes the coupling energy between two anti-parallel sublattice skyrmions, i.e., the coupling energy of a resting AFM skyrmion $W_c^0$.  The second term vanishes because of Eq.\ \eqref{eq:inter_integrals}, while the third term is $\frac{1}{2}  \sum_{\mu} \delta_{\mu}^2 \Lambda$. Overall, this leads to an energy
 \begin{equation}
 	\label{eq:energy_disp}
 	W_{\text{total}} = W_0^A + W_0^B + W_c^0 + \frac{1}{2} \Lambda |\bm{\delta}|^2,
 \end{equation}
which is the energy of a resting AFM skyrmion plus the excitation energy $\Delta W = \Lambda|\bm{\delta}|^2 / 2$. 
For a free moving AFM skyrmion, we use that  $\bm{v} = \Lambda \bm{\delta} / (4 \pi)$  from Eq.\ \eqref{eq:Thiele_A_B_damp}, as well as the effective mass $m = 16 \pi^2 / \Lambda$, $\Delta W$ can be interpreted as the kinetic energy $E_{\text{kin}} = \frac{m}{2} |\bm{v}|^2$ of a free particle. Thus, each (massless) FM sublattice skyrmion experiences the force $\bm{F} = - \Lambda \bm{\delta}$, see Eqs.\ \eqref{eq:force_AB} and \eqref{eq:force_BA}, which stems from the harmonic potential created by the respective other sublattice. This force acts {\em internally}, drives the FM sublattice skyrmions to a motion with constant velocity, in the spirit of Thiele's equation (cf.\ Ref.\ \cite{DispPanigrahy}), and creates the AFM skyrmion inertia. {\em External} forces, that accelerate the skyrmion, then cause the AFM skyrmion to move like a classical, massive, and damped particle. Possible excitations of internal skyrmion modes are by construction not included in the Thiele approach, but might show up in numerical simulations of the full skyrmion dynamics (see below). 
	
\section{Current driven skyrmions}
 \label{sec:sk_velo}
	For current-driven skyrmions \cite{Parkin2008}, we assume that spin polarized electrons travel through either sublattice \cite{BarTre2016} and exert torques on the magnetic moments of the respective FM sublattice \cite{GomoCurAFM}. For each sublattice, $\bm{n} \in \{\bm{a},\bm{b}\}$, the LLG generalizes to $\partial_t \bm{n} = -\bm{n} \times \bm{H}_{\text{eff}} + \alpha \bm{n} \times \partial_t \bm{n} + (\bm{j} \bm{\nabla}) \bm{n} - \beta \bm{n} \times (\bm{j} \bm{\nabla}) \bm{n}$ \cite{StierPRL17}, as for a bare FM, where $\bm{j}$ denotes the current density, including the degree of spin polarization as a prefactor. It contains also the nonadiabatic spin torque of strength $\beta$ \cite{ModSTT}. Although in most cases the currents may flow in parallel in both sublattices \cite{BarTre2016,HalsTserkBrat,Velkov}, here we allow the currents $\bm{j}_{A/B}$  to flow in different directions. This may seem difficult to achieve in an AFM, however, it can be realized in antiferromagnetic coupled bilayer systems \cite{KoshBilayer}. In general, the LLG equations for the two sublattices
	\begin{equation}
		\label{eq:derive1}
		\begin{aligned}
			- (\bm{v}\bm{\nabla}) \bm{a} = &-\bm{a} \times \bm{H}_{\text{eff}}^A - \alpha \bm{a} \times (\bm{v}\bm{\nabla}) \bm{a}\\ 
			&+ (\bm{j}_A \bm{\nabla}) \bm{a} - \beta \bm{a} \times (\bm{j}_A \bm{\nabla}) \bm{a}\, ,
		\end{aligned}
	\end{equation}
	\begin{equation}
	\label{eq:derive2}
	\begin{aligned}
		- (\bm{v}\bm{\nabla}) \bm{b} = &-\bm{b} \times \bm{H}_{\text{eff}}^B - \alpha \bm{b} \times (\bm{v}\bm{\nabla}) \bm{b}\\ 
		&+ (\bm{j}_B \bm{\nabla}) \bm{b} - \beta \bm{b} \times (\bm{j}_B \bm{\nabla}) \bm{b}
	\end{aligned}
\end{equation}
can be brought in the form of Thiele's equation, similiar to Eq.\ \eqref{eq:thiele_sk_sl}.  This results in
\begin{equation}
	\label{eq:thiele_sk_sl_current}
	\begin{aligned}
		-\bm{G} \times (\bm{v}_A + \bm{j}_A) - \alpha \mathcal{D} \bm{v}_A   - \beta \mathcal{D} \bm{j}_A- \Lambda \bm{\delta} &= 0 \, , \\
		\bm{G} \times (\bm{v}_B + \bm{j}_B) - \alpha \mathcal{D} \bm{v}_B - \beta \mathcal{D} \bm{j}_B + \Lambda \bm{\delta} &= 0\, .
	\end{aligned}
\end{equation}
We note that shifting all terms of Eq.\ \eqref{eq:thiele_sk_sl_current} to the right-hand-side of the equation would result in two coupled Thiele's equations with external forces as considered in Ref.\ \cite{DispPanigrahy}. By identifying $\partial_t \bm{\delta} = \bm{v}_A - \bm{v}_B$ and $\bm{v} = (\bm{v}_A + \bm{v}_B)/2$ as center-of-mass velocity of the AFM skyrmion, we construct an equation of motion for the displacement as 
	\begin{equation}
		\label{eq:ode_disp}
		\partial_t \bm{\delta} + \frac{\alpha \mathcal{D} \Lambda}{8 \pi^2} \bm{\delta} + \bm{j}^{-} + \frac{(\alpha - \beta) \mathcal{D}}{4 \pi Q} \bm{z} \times \bm{j}^{+}=0 \, ,
	\end{equation}
	with the effective current densities $\bm{j}^{-} = \bm{j}_A - \bm{j}_B$ and $\bm{j}^{+} = \bm{j}_A + \bm{j}_B$. Solving Eq. \eqref{eq:ode_disp} with an initial value $\delta_0 = 0$ yields the time-dependent displacement
	\begin{equation}
	\label{eq:sol_ode_disp}
	\begin{aligned}
	\bm{\delta}(t) &= \frac{8 \pi^2}{\alpha \mathcal{D} \Lambda} \left(\exp\left[\frac{-\alpha \mathcal{D} \Lambda}{8 \pi^2} t\right] - 1\right) \bm{j}^{-}\\ 
	&+ \frac{2 \pi (\alpha - \beta)}{\alpha \Lambda Q} \left(\exp\left[\frac{-\alpha \mathcal{D} \Lambda}{8 \pi^2} t\right] - 1\right)\bm{z} \times \bm{j}^{+}.
\end{aligned}
	\end{equation}
	The total skyrmion velocity then follows as
	\begin{equation}
		\label{eq:final_sk_motion}
		\bm{v} = \frac{\Lambda}{4 \pi Q} \bm{z} \times \bm{\delta} - \frac{1}{2} \bm{j}^{+} - \frac{(\alpha - \beta) \mathcal{D}}{8 \pi Q} \bm{z} \times \bm{j}^{-} \, .
	\end{equation} 
	We examine two scenarios, in both of which we set  $|\bm{j}_A|=|\bm{j}_B|=j$. In the first scenario, we consider unidirectional  sublattice currents, $\bm{j}^{+}=2 \bm{j}$ and $\bm{j}^{-}=0$, in the second scenario two currents flow in opposite directions, $\bm{j}^{+}=0$ and $\bm{j}^{-}=2 \bm{j}$. For both, the  equation of motion, Eq.\ \eqref{eq:ode_disp}, can be solved for constant currents and an initially undisplaced skyrmion, $\bm{\delta} (t=0) = 0$. The resulting time-dependent displacement $\bm{\delta} (t)$ can then be inserted into Eq.\ \eqref{eq:final_sk_motion}.
	
	\subsection{Unidirectional currents}
	This case has already been studied for both single layer \cite{BarTre2016} and synthetic \cite{BilayerSk} AFM skyrmions, however, without taking the full time evolution for the displacement between the sublattice skyrmions into account. The time dependent AFM skyrmion moves in the direction of the current with a velocity
	\begin{equation}
		\label{eq:sk_motion_current_para}
		v_\parallel(t) = -j \left( \frac{\beta}{\alpha} + \frac{\alpha - \beta}{\alpha} \exp\left[-\frac{\Gamma \Lambda}{8 \pi^2}t\right] \right)\, .
	\end{equation}
	FM skyrmions drift along with the current flow and exhibit a skyrmion Hall effect (SHE) \cite{NagaTokOverview} whose sign depends on the sign of the topological charge $Q$ \cite{StierPRL17}. The two sublattice skyrmions exhibit opposite topological charges $Q_A = -Q_B$ \cite{BarTre2016, ZhangAFMSk}. Thus, their individual SHEs occur in opposite directions, leading to an overall vanishing SHE of the AFM skyrmion. Yet, this influences the sublattice skyrmion displacement and therefore the AFM skyrmion velocity. Equation \eqref{eq:sk_motion_current_para} recovers the asymptotic velocity $v=\frac{\beta}{\alpha} j$ of Ref.\ \cite{BarTre2016} and additionally provides an analytical solution at arbitrary times. For  $\alpha = \beta$, the FM sublattice skyrmions show no SHE \cite{StierPRL17} and the AFM skyrmion velocity equals that of the current driven FM skyrmion.

	\subsection{Antiparallel sublattice currents} 
	The second scenario leads to an AFM skyrmion motion perpendicular to the current direction, with the velocity
	\begin{equation}
		\label{eq:sk_motion_current_perp}
		v_\perp (t) = j \frac{4 \pi Q}{\Gamma} \left( 1 - \exp\left[-\frac{\Gamma \Lambda}{8 \pi^2}t\right] \right).
	\end{equation}
	The applied current pushes the two sublattice skyrmions into opposite directions and, thus, increases the displacement between them which, in turn, drives a velocity. We emphasize that magnitude and direction of this velocity is almost entirely determined by the displacement and {\em not} by the SHE on each sublattice. Therefore, no parameter $\beta$ appears in Eq.\ \eqref{eq:sk_motion_current_perp} The asymptotic velocity $v_\perp = 4 \pi Q j/ \Gamma $ depends inversely on the damping and coincides with that of Ref.\ \cite{KoshBilayer}.
	
	While the asymptotic velocity only depends on the skyrmion parameters (via $\mathcal{D}$), the time needed to reach it depends on $\Lambda$. Hence, in a bilayer synthetic AFM the inter-layer coupling $J_{\text{inter}}$  determines this time constant. The sign of the velocity in Eq.\ \eqref{eq:sk_motion_current_perp} depends on $Q$, since we chose sublattice A as reference for both the skyrmion as well as the current. Revisiting the picture of a classical particle, Eq.\ \eqref{eq:sk_motion_current_perp} is the solution to  $m \ddot{x} + \Gamma \dot{x} = -4 \pi Q j$ of an initially resting classical particle driven by an external force of strength $4 \pi Q j$.
	
	\section{Comparison to numerical simulations}
	\label{sec:simulation}
    To verify the model predictions, we compare them to simulations. For a single layer AFM on a discrete two-dimensional square lattice of lattice constant $d$, we numerically solve the LLG for each lattice site. We use  a lattice Hamiltonian, similar to Ref. \cite{ZhangAFMSk}, in the form 
    \begin{equation}
    	\label{eq:lattice_hamiltonian}
    	\begin{aligned}
    		H =&\ \tilde{J} \sum_{\bm{r}} \bm{n}_{\bm{r}} \cdot \left(\bm{n}_{\bm{r} + \bm{x}} +\bm{n}_{\bm{r} + \bm{y}} \right)\\ 
    		&+ \tilde{D} \sum_{\bm{r}} \left( \left[\bm{n}_{\bm{r}} \times \bm{n}_{\bm{r} + \bm{x}} \right] \cdot \bm{x} + \left[\bm{n}_{\bm{r}} \times \bm{n}_{\bm{r} + \bm{y}} \right] \cdot \bm{y} \right)\\ 
    		&+ \tilde{K} \sum_{\bm{r}} n_{\bm{r}}^{z} \, ,
    	\end{aligned}
    \end{equation}
   and set $\tilde{D}/\tilde{J} = 0.045$ and $\tilde{K}/\tilde{J} = 0.013$ to stabilize a stationary skyrmion. Since the parameters from the field theory \eqref{eq:conti_energy_afm} are not the same, we mark the parameters of the lattice model by a tilde. In the limit of small lattice constants, the energy given in Eq.\ \eqref{eq:conti_energy_afm} resembles \eqref{eq:lattice_hamiltonian} by putting $A^{\prime}$, $D^{\prime}$, and $K$ to zero. The size of the stabilized skyrmion used here, according to a fit to the skyrmion profile in Eq.\ \eqref{eq:fit_theta}, assumes then the radius $R/d = 13.3$ and the width $w/d = 5.6$. The time scale is in units of $t_0=1/\tilde{J}$.
	
\subsection{Free motion}
To simulate the undriven skyrmion dynamics, we initially displace the two sublattice skyrmions by $\bm{\delta}_0$, which is less than one lattice site, $|\bm{\delta}| \ll d$, using an approximated analytical expression for the skyrmion. Imposing such a small displacement on a lattice poses a nontrivial task. To this end, we substitute the numerical components $(n_x(x,y), n_y(x,y), n_z(x,y))$ of the magnetic moments of a stationary AFM skyrmion on lattice points $(x,y)$ by $\pm(- \sin\varphi\:\sin\theta(\rho),\cos\varphi\:\sin\theta(\rho),\cos\theta(\rho))$ for the two sublattices A and B, respectively, where $\rho=\sqrt{x^2+y^2}$. Since the skyrmion profile $\theta (\rho)$ cannot be expressed by known analytical functions we use the approximation of Ref.\ \cite{SkySize}
		\begin{equation}
			\label{eq:fit_theta}
			\theta(\rho) = 2 \arctan \left( \frac{\sinh \left[R/w\right]}{\sinh \left[\rho/w\right]} \right) \, , 
		\end{equation}
		which turns out to work very well in all cases investigated here. Fitting the FM skyrmion on each sublattice to Eq.\ \eqref{eq:fit_theta} confirms that both sublattice skyrmions indeed have the same profile. By the fitting, we obtain the radius $R$ and the width $w$ of the skyrmion. Then, we repeat the step for the other sublattice with the same expression, however, using $\bm{r} + \bm{\delta}$ as skyrmion coordinates and transform it back to its originally used lattice representation. Recombining the two sublattices gives an AFM skyrmion with a sublattice displacement $\bm{\delta}$. \\
		
\begin{figure}[t!]
			\includegraphics[width=0.5\textwidth]{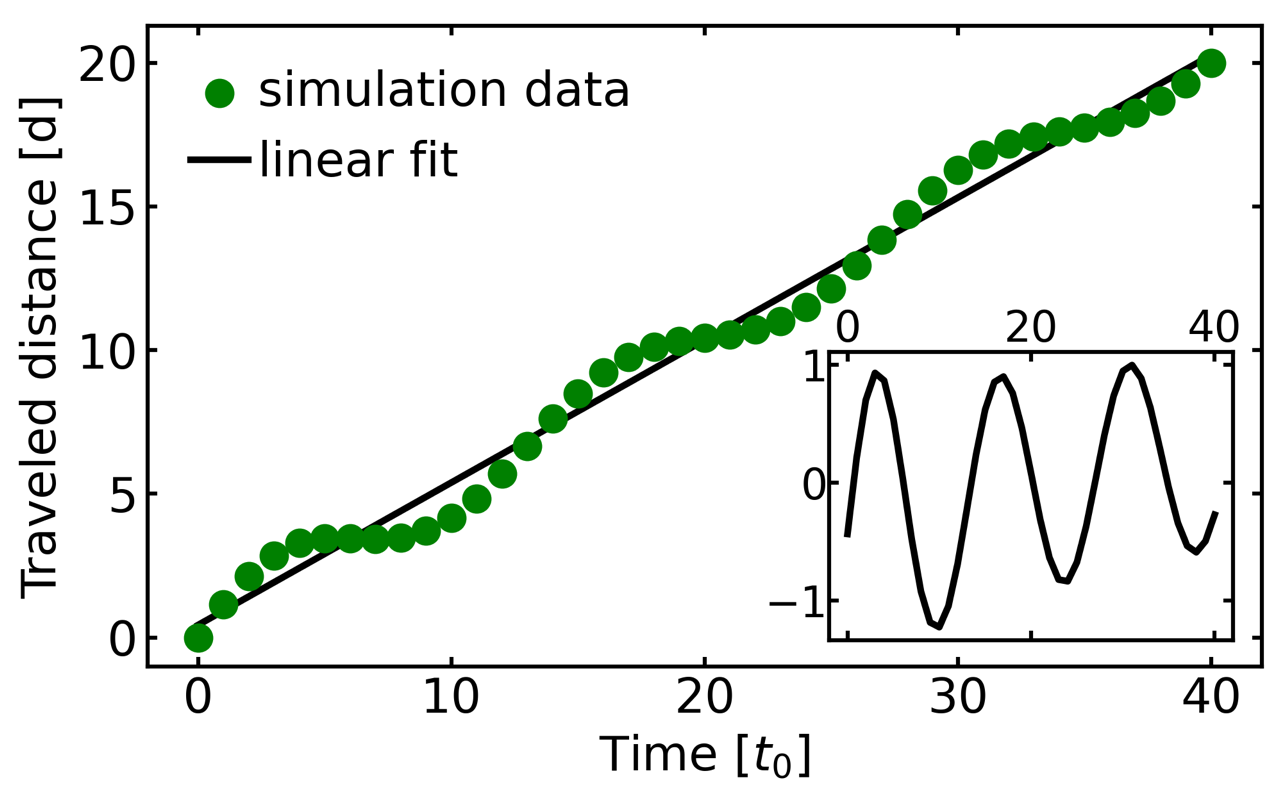}
			\caption{a) Time-dependent distance traveled by an AFM skyrmion with an initial displacement $\delta_0$ (main). The green dots mark the distance according to the numerical simulations, while the straight solid line is a linear fit. The actual skyrmion motion oscillates around the uniform motion with a constant velocity. The inset illustrates these oscillations by depicting the difference between the full distance traveled and the linear fit. 
			\label{fig:art_disp_travel}}
	\end{figure}

		As predicted in Sec.\ \ref{sec:sk_velo}, the initial displacement leads to a skyrmion motion with a constant velocity perpendicular to the displacement direction for the non-damped case $\alpha=0$. For rigid skyrmions a uniform motion with constant velocity would result as the only motion. However, the simulation on the basis of the Landau-Lifshitz-Gilbert equation for each magnetic moment reveals also inner excitation modes and deformations of the skyrmion at finite displacements. They explain the oscillatory motion seen in Fig.\ \ref{fig:art_disp_travel}, where the distance traveled by the skyrmion is shown after the initial displacement. The dots are the skyrmion position over time while the solid line is a linear fit. The inset shows the difference between the real motion and the linear fit,	highlighting the oscillations. We only find oscillations in the  free skyrmion motion induced by the initial sublattice displacement. On the contrary, no oscillations are found after driving initially undisplaced skyrmions with the electric current. We therefore conclude that the oscillation stems from the instantaneous excitation of the skyrmion due to the artificially created initial displacement.
		
\begin{figure}[t!]
		\includegraphics[width=0.48\textwidth]{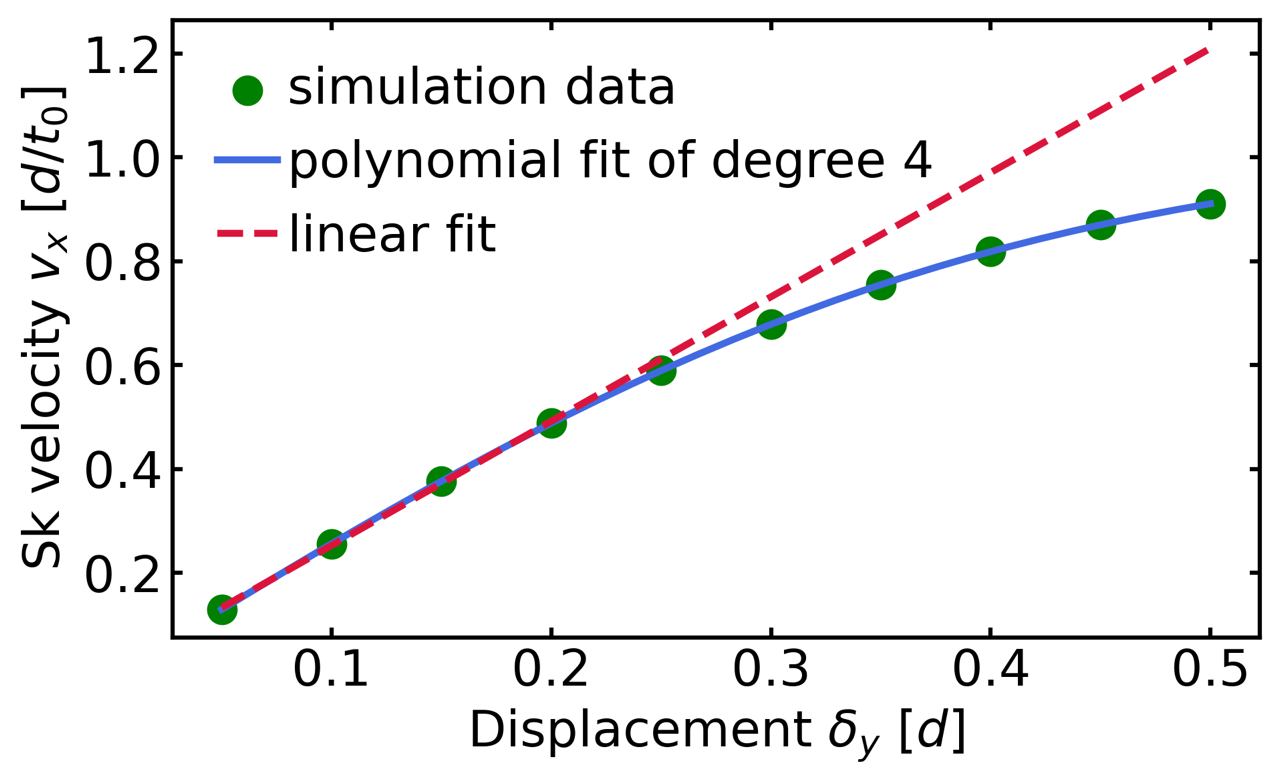}
		\caption{Velocity of the AFM skyrmion for different initial displacements $\delta_0=\delta_y$. The green dots represent data extracted from numerical simulations. The dashed red line is a linear fit to the lowest four data points, the solid blue line is a fit to  $v(\delta) = 4.4 \delta^4 -5.8 \delta^3 + 0.3 \delta^2 + 2.6 \delta$. We find that Eq.\ \eqref{eq:v_AFM_SK_damp} correctly predicts the velocity for small displacements $\delta_y \lesssim 0.2 d$, here for $\alpha=0$.
			\label{fig:v_over_delta_x0}}
	\end{figure}

	We have performed simulations for several initial displacements $\delta_y$ and have extracted the mean skyrmion velocity by a linear fit. The results are shown in Fig.\ \ref{fig:v_over_delta_x0} versus the initial displacement as green dots. The solid blue line is a fit to a fourth order polynomial, while a linear dependence, according to Eq.\ \eqref{eq:v_AFM_SK_damp}, accounts correctly for small displacements $\delta_y \lesssim 0.2 d$ (dashed red line). For larger values of $\delta_y$, the fit yields $v(\delta) = 4.4 \delta^4 -5.8 \delta^3 + 0.3 \delta^2 + 2.6 \delta$, suggesting that our formalism could be extended by including higher-order terms in $\bm{\delta}$.
	
\begin{figure}[t!]
		\includegraphics[width=0.5\textwidth]{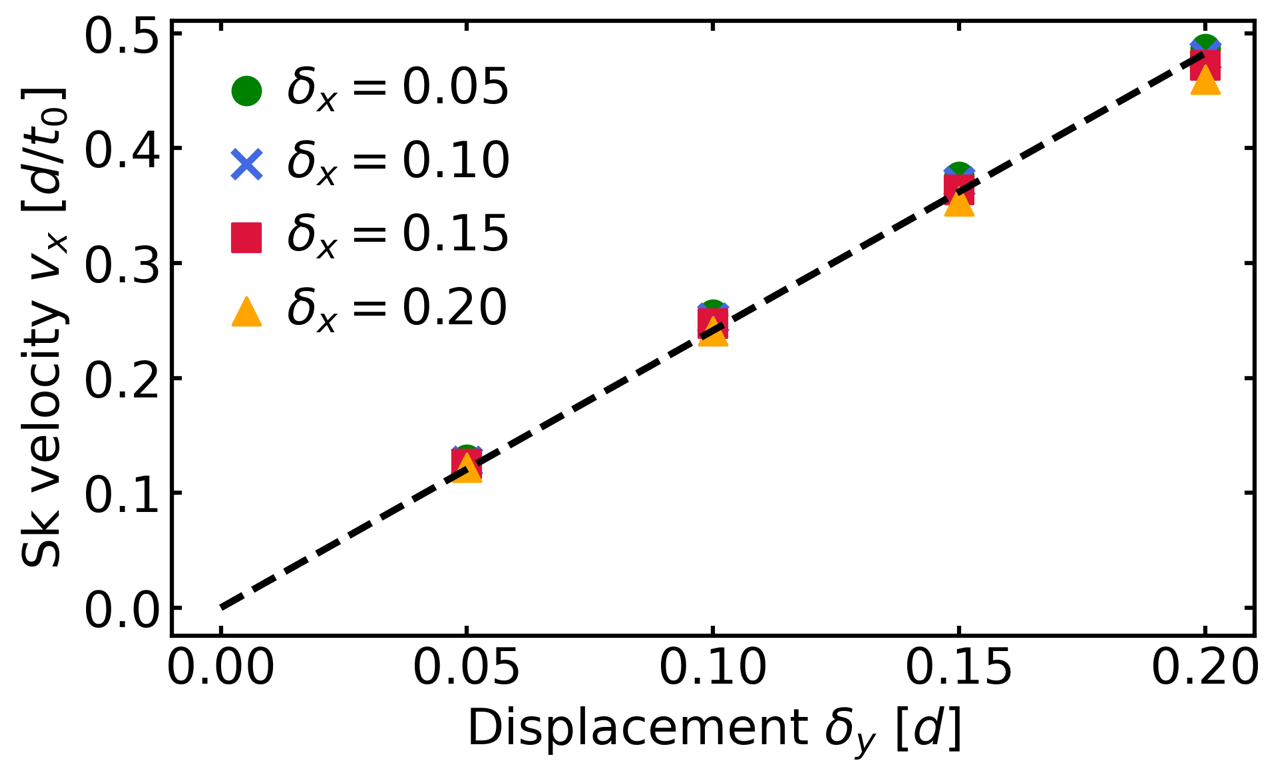}
		\caption{Velocity of the AFM skyrmion in $x$-direction plotted against the displacement $\delta_y$ considered in the simulations. The four different symbols (and colors) belong to  different values of the  initial displacements $\delta_x$ in $x$-direction. Since one can barely distinguish these different data, we can conclude that  $v_x$ is independent from the displacement in $x$-direction, particularly in the regime of small $\delta_y$.
			\label{fig:art_disp_dif_del}}
	\end{figure}

\begin{figure}[t!]
		\includegraphics[width=0.48\textwidth]{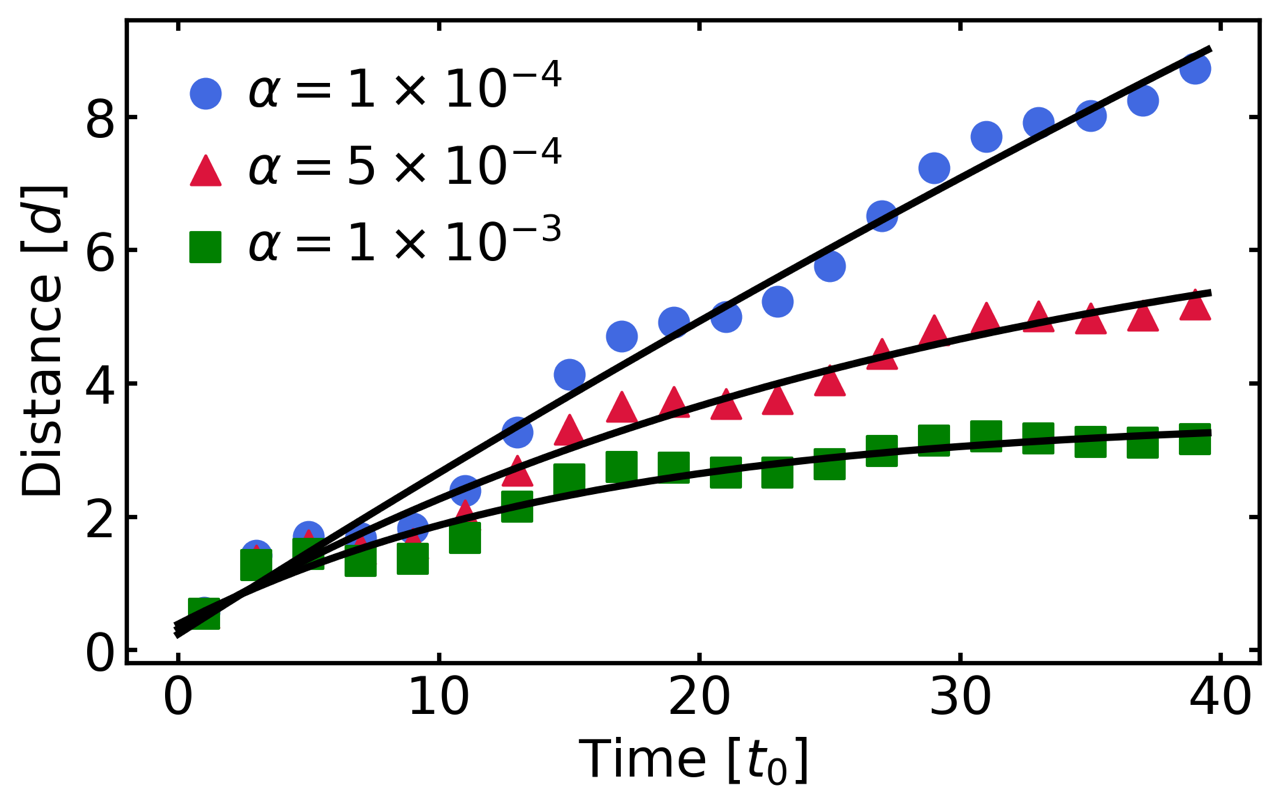}
		\caption{Distance traveled by the AFM skyrmion along the lattice over time, starting from an initial displacement $\delta_0= 0.1 d$. The symbols mark the simulation results for different values of the damping $\alpha$ as indicated, while the solid lines show fits to Eq.\ \eqref{eq:v_AFM_SK_damp}.
			\label{fig:damped_distance_over_time}}
	\end{figure}

	To confirm that the velocity $v_x$ in $x$-direction is indeed independent of the displacement $\delta_x$, we also depict $v_x$ as a function of $\delta_y$ for different values of $\delta_x$ in Fig.\ \ref{fig:art_disp_dif_del}. Indeed, the skyrmion velocity only depends on displacmenet components perpendicular to it, in agreement with Eq.\ \eqref{eq:Thiele_A_B_damp}.
	For finite damping $\alpha > 0$, the skyrmion motion over time is depicted in Fig.\ \ref{fig:damped_distance_over_time}. Symbols show the distance traveled, while solid lines are fits using Eq.\ \eqref{eq:v_AFM_SK_damp}. As for $\alpha = 0$, the simulations data exhibit additional oscillations of amplitudes that are small compared to the entire path length. They result from inner mode excitations accompanied with the displacement in each of the FM constituents, and are not included in the analytical treatment which assumes rigid sublattice skyrmion profiles.

	\subsection{Current driven skyrmion}
	Driving an initially resting AFM skyrmion by a spin-polarized current, similar to Ref. \cite{StierPRL17}, allows us to extract the AFM skyrmion velocity for both the scenarios mentioned in Sec.\ \ref{sec:sk_velo}.
	\begin{figure}[t!]
		\includegraphics[width=0.48\textwidth]{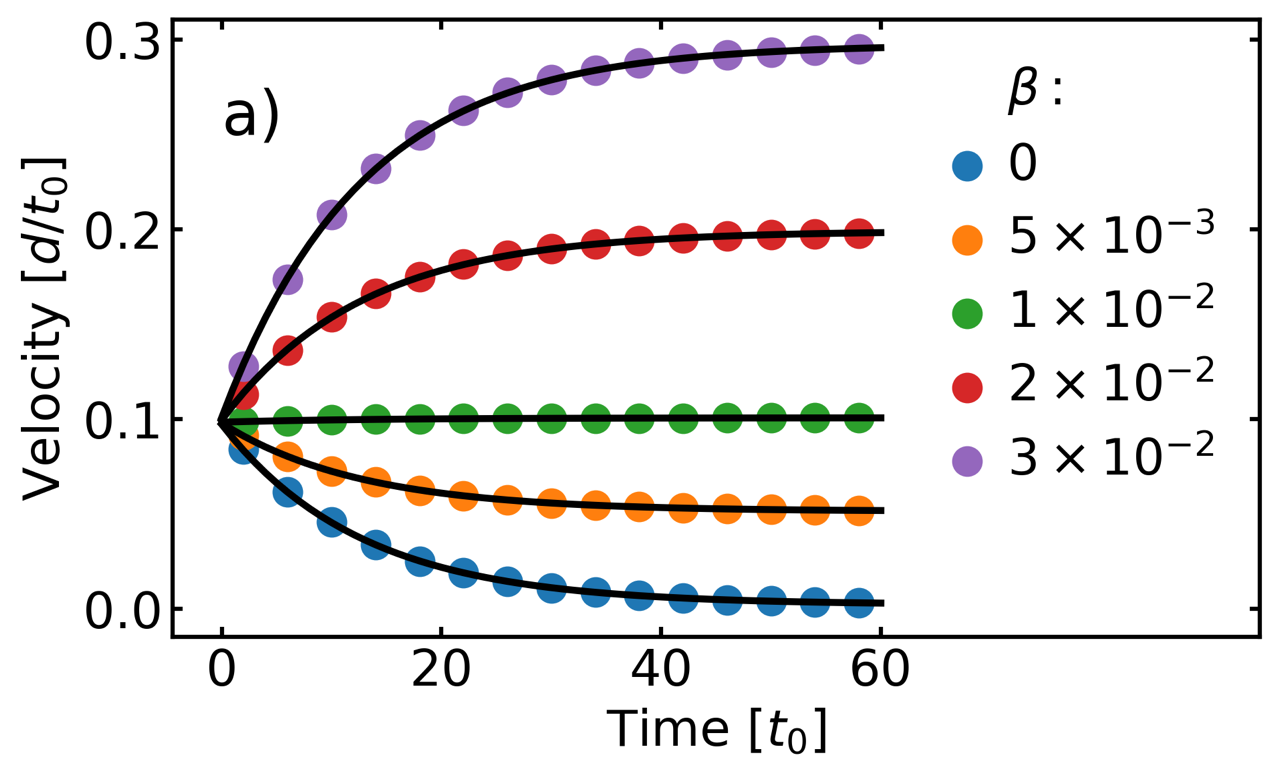}
		\includegraphics[width=0.48\textwidth]{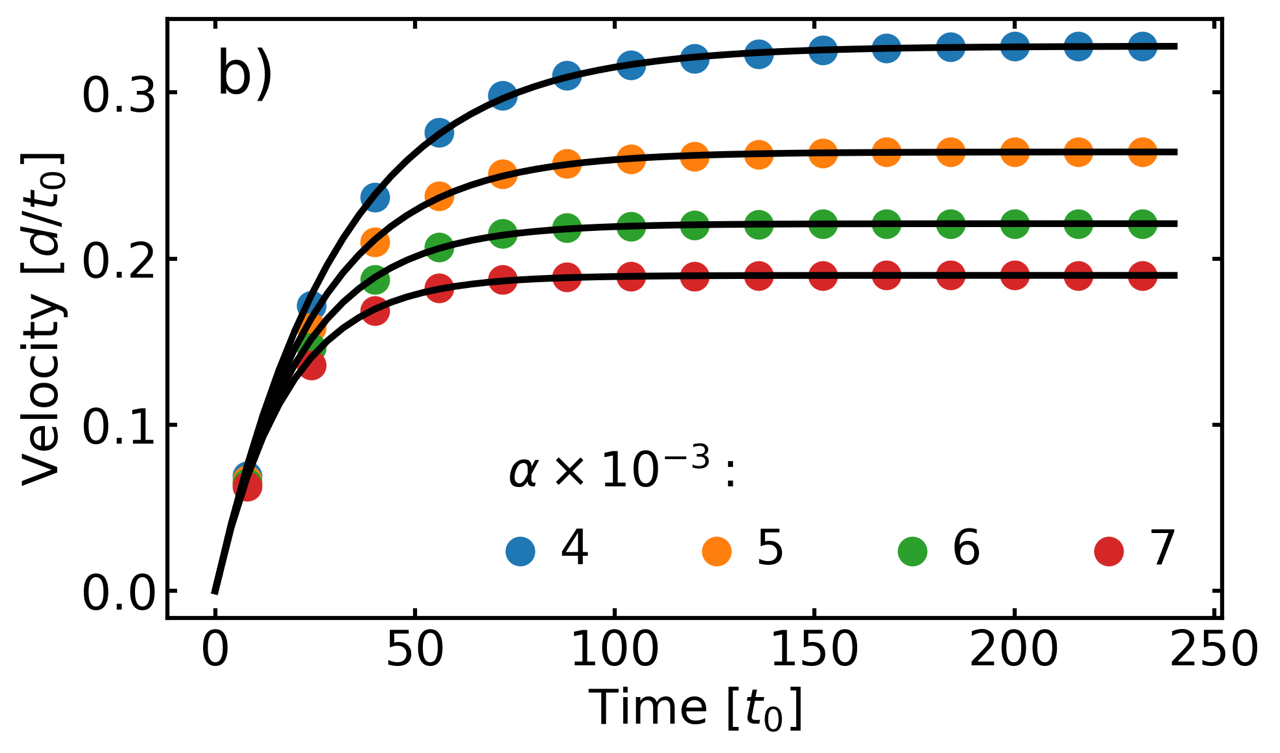}
		\caption{Velocity of a current-driven AFM skyrmion over time. The symbols show the results extracted from the simulation data, while the solid lines are obtained from Eqs.\ \eqref{eq:sk_motion_current_para} (a) and \eqref{eq:sk_motion_current_perp} (b), respectively. (a) Unidirectionally applied currents with $\bm{j}_A = \bm{j}_B = 0.1\ d/t_0$ for different values of $\beta$ for $\alpha = 10^{-2}$. (b) Antiparallel applied currents with $\bm{j}_A = -\bm{j}_B = 0.002\ d/t_0$ for different values of $\alpha$ for $\beta=10^{-2}$.
			\label{fig:current_driven}}
	\end{figure}
	Fig.\ \ref{fig:current_driven} (a) shows the velocities for different values of the nonadiabaticity parameter $\beta$ for fixed damping $\alpha = 0.01$ and an applied current corresponding to $\bm{j}_{A/B} = 0.1\ d / t_0$. The solid lines result from a fit of Eq.\ \eqref{eq:sk_motion_current_para} to the simulation data. The optimal fit yields $\mathcal{D} / m \approx 7.8 J$. The scenario studied in Fig.\ \ref{fig:current_driven} (b) pushes sublattice skyrmions apart and, thus, induces relatively large velocities. We therefore chose a low current density of $\bm{j}_{A/B} = 0.002\ d / t_0$. Figure \ref{fig:current_driven} (b) shows velocities for different values of $\alpha$ and fixed $\beta = 0.01$, all in the order of $\mathcal{O}(10^{-3})$. Solid lines follow from a fit to Eq.\ \eqref{eq:sk_motion_current_perp} which yields the parameters $m \approx 2.4 / J$ and $\mathcal{D} \approx 19$ separately. They coincide with the fitted parameters from Fig.\ \ref{fig:current_driven} (a).

	Finally, we mention that the value of $\Lambda$ obtained from the fitting to the simulation data differs from  the value obtained by numerically solving the integral from Eq.\ \eqref{eq:lambda_def}   by roughly a factor of 2. Similar to the oscillatory movement seen in Fig.\ \ref{fig:damped_distance_over_time} we attribute this difference to excitations of inner modes, deforming the FM skyrmion constituents while being dragged apart, already to linear order in $\bm{\delta}$. Such a deformation arises naturally due to the coupling between FM skyrmion constituents.  On the other hand, deformations of sublattice constituents are not included in Thiele's approach. According to our findings, only the fitted value of $\Lambda$ is affected, while, for example, the dissipation constant $\bm{\mathcal{D}}$ is found to agree quantitatively with the results of the numerical simulations. Inclusion of inner skyrmion modes to the theory is postponed to future research.
	
	\subsection{Skyrmion motion on a synthetic bilayer AFM}
\begin{figure}[t!]
		\includegraphics[width=0.48\textwidth]{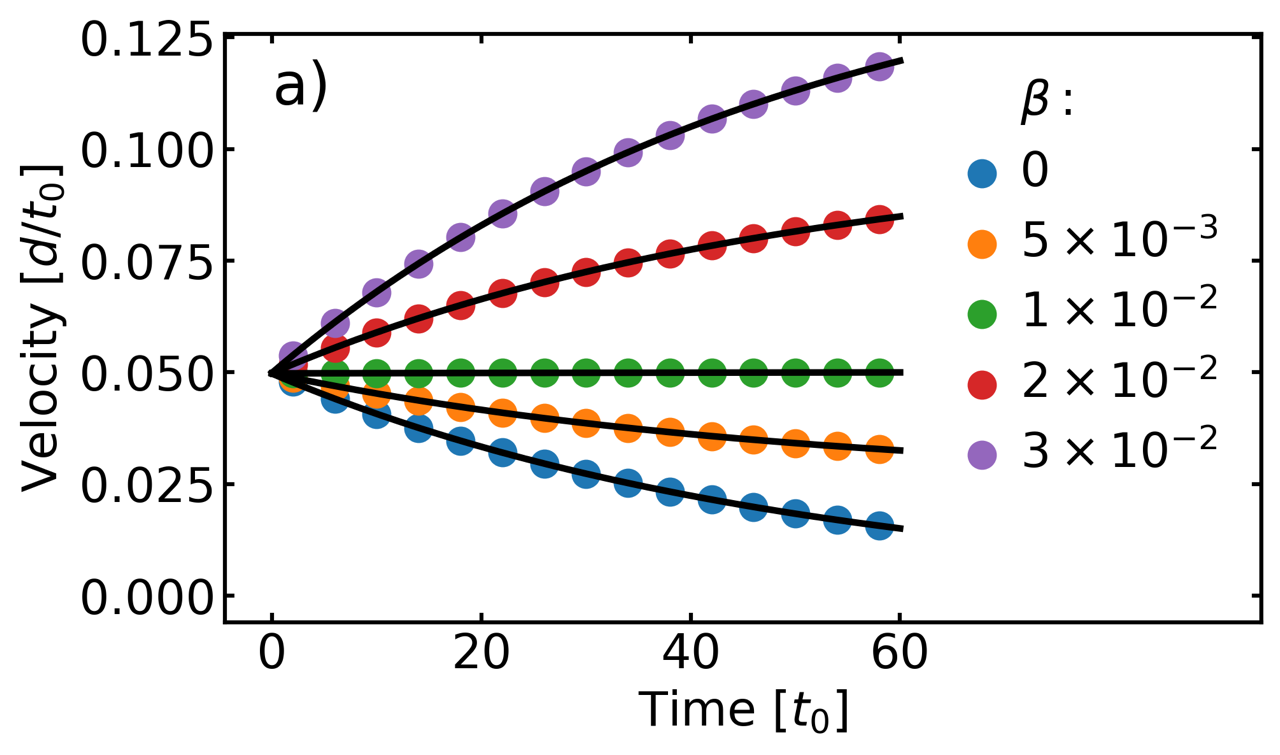}
		\includegraphics[width=0.48\textwidth]{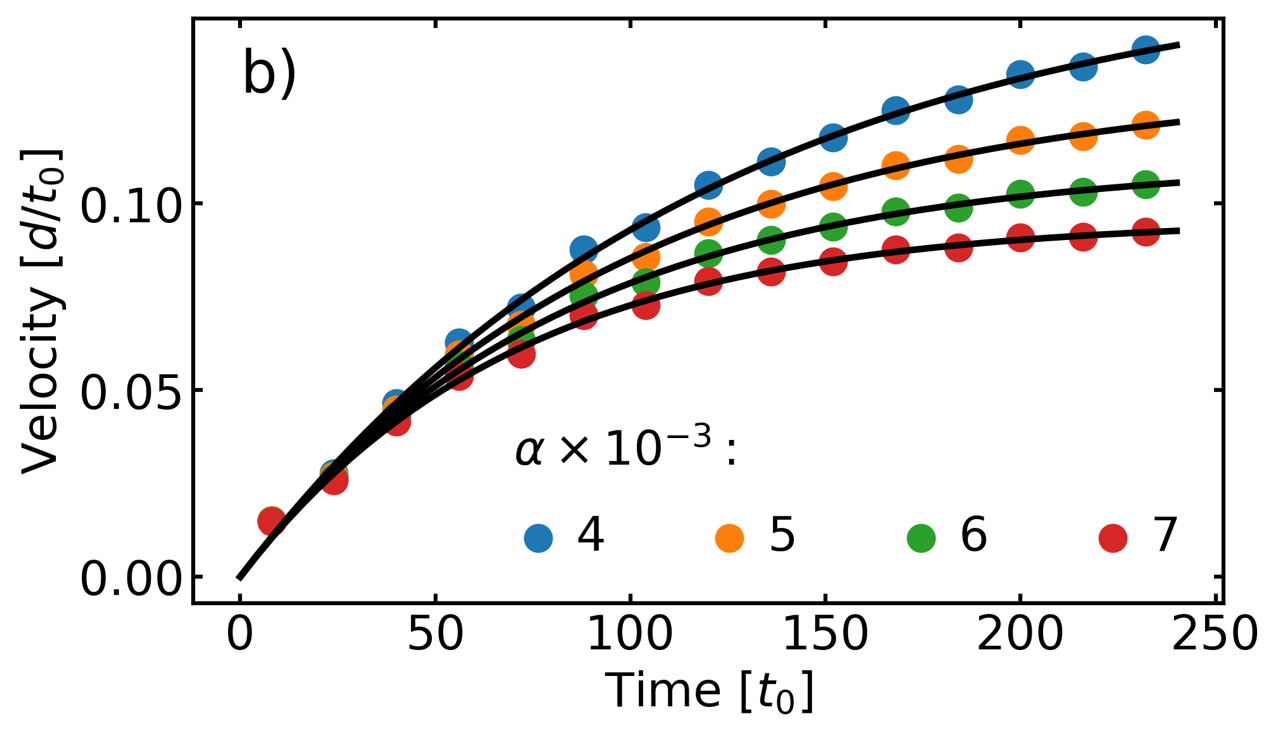}
		\caption{Velocity of a current-driven synthetic AFM skyrmion over time. The symbols show the results extracted  from the simulation data, while the solid lines are obtained from Eqs.\ \eqref{eq:sk_motion_current_para} and \eqref{eq:sk_motion_current_perp} adapted to the case of a synthetic AFM. a) Parallel (unidirectionally) applied currents with $\bm{j}_A = \bm{j}_B = 0.05\ d/t_0$ for different values of $\beta$ for $\alpha = 10^{-2}$. b) Antiparallel applied currents with $\bm{j}_A = -\bm{j}_B = 0.001\ d/t_0$ for different values of $\alpha$ for $\beta=10^{-2}$.
			\label{fig:cur_syn}}
	\end{figure}

	All simulations shown in the previous subsections are carried out on a discrete single layer AFM lattice as in Ref.\ \cite{SkSw}. We consider a two-dimensional square lattice with nearest neighbor interactions composed of two sublattices ordered in the checkerboard arrangement. Since our formalism remains qualitatively unchanged (with only the parameter $\Lambda$ to be adapted) for different kinds of antiferromagnets, we have additionally simulated a synthetic bilayer AFM, as another interesting example, also on a discrete square lattice. We consider two ferromagnetic sublattices with an antiferromagnetic on-site coupling, $(\lambda/2)\,\bm{a} \cdot \bm{b}$, as the only coupling term between sublattices, cf.\ Refs.\ \cite{KoshBilayer,DispPanigrahy} and Sec.\ \ref{sec:sk_velo}. The lattice Hamiltonian used for the simulation is, similar to Ref.\ \cite{KoshBilayer},
	\begin{equation}
		\label{eq:lattice_hamiltonian_bilayer}
		\begin{aligned}
		H =&\ H_{\text{FM}}^A + H_{\text{FM}}^B - \tilde{J}_{\text{inter}} \sum_{\bm{r}} \bm{n}_{\bm{r}}^A \cdot \bm{n}_{\bm{r}}^B,\\
		H_{\text{FM}}^{i} = &-\tilde{J} \sum_{\bm{r}} \bm{n}_{\bm{r}}^{i} \cdot \left(\bm{n}_{\bm{r} + \bm{x}}^{i} +\bm{n}_{\bm{r} + \bm{y}}^{i} \right)\\ 
		&+ \tilde{D} \sum_{\bm{r}} \left( \left[\bm{n}_{\bm{r}}^{i} \times \bm{n}_{\bm{r} + \bm{x}}^{i} \right] \cdot \bm{x} + \left[\bm{n}_{\bm{r}}^{i} \times \bm{n}_{\bm{r} + \bm{y}}^{i} \right] \cdot \bm{y} \right)\\ 
		&+ \tilde{K} \sum_{\bm{r}} n_{\bm{r}}^{i,z}.
	\end{aligned}
\end{equation}
In the limit of small lattice constants, it is associated to the energy given in  Eq.\ \eqref{eq:conti_energy_afm} by puting the parameters $A$, $D$ and $K$ to zero, and by identifying $\frac{A^{\prime}}{2} \rightarrow \frac{\tilde{J}}{2}$, $\frac{D^{\prime}}{2} \rightarrow \frac{2\tilde{D}}{d}$, $\frac{K^{\prime}}{2} \rightarrow  \frac{\tilde{K}}{d^2}$, and, as the only coupling between the sublattices, $\frac{\lambda}{2} \rightarrow \frac{\tilde{J}_{\text{inter}}}{d^2}$. In our simulations we assume equal interlayer and exchange coupling strength $\tilde{J}_{\text{inter}} = |\tilde{J}|$, i.e., for the continuum approximation $\lambda/2 = A'$. This leads to a force factor of $\Lambda = |\tilde{J}| \mathcal{D}/d^2$ in the simulations. In Fig.\ \ref{fig:cur_syn}, the current driven skyrmion motion in the bilayer model is shown. Regarding the resulting skyrmion velocity, we refer to Ref.\ \cite{KoshBilayer} where a similar model was simulated. An estimate of the time unit of $t_0 \approx 0.7$ ps and of a lattice constant $d = 5\ \mathring {\mathrm A}$ is reported. Thus units for skyrmion velocities are converted by $1 \frac{d}{t_0} \approx 714 \frac{m}{s}$. Using also the same representation of the current induced torque as Ref.\ \cite{KoshBilayer}, the unit of the electric current would lie, accordingly, in the order of magnitude $\sim 5 \times 10^{11}$ A/m$^2$ for the case of uni-directional currents and $\sim 1 \times 10^{10}$ A/m$^2$ for the case of anti-parallel flowing currents. We here consider only slowly moving skyrmions in order to stay as close as possible to the rigid skyrmion approximation. We find a qualitatively similar motion as for the skyrmion in a single layer AFM (see Fig.\ \ref{fig:current_driven}). Here, the inertial mass is four times larger as compared to Fig.\ \ref{fig:current_driven}, which stems from the fact that in the single layer lattice each sublattice spin has four neighbors on the respective other sublattice, as compared to only one in the bilayer case. This directly affects the respective effective field strengths and thereby $\Lambda$. The skyrmion moves along the current direction if the sublattice currents flow in the same direction, $\bm{j}_A = \bm{j}_B$ (Fig.\ \ref{fig:cur_syn} a)), while it moves perpendicular to the current for anti-parallel sublattice currents, $\bm{j}_A = - \bm{j}_B$ (Fig.\ \ref{fig:cur_syn} b)).
    For the bilayer synthetic AFM case the positions of the two FM sublattice skyrmion constituents are particularly easily extracted \cite{StierStro}. We find indeed a displacement $\bm{\delta}$ proportional to  $\bm{v} + \bm{j}$ for unidirectional currents, and to $\bm{v}$ for anti-parallel currents, consistent with Eq.\ \eqref{eq:final_sk_motion}. The extracted value of $\Lambda \approx 8.6$ from fitting the skyrmion velocity to these equations, however, differs from the predicted value $\Lambda = |\tilde{J}| \mathcal{D} / d^2 = 18.5$, where we calculated $\mathcal{D}$ by assuming a skyrmion profile as in Eq.\ \eqref{eq:fit_theta} and setting $d$ and $\tilde{J}$ to one. Simulations describe the full dynamics of the vector field without the rigid-skyrmion approximation. They include inner excitation modes and skyrmion deformation. We find that those excitations, especially the deformation, lead to a different value of $\Lambda$ compared to the one predicted by the Thiele picture. They are also the reason for the oscillations apparent in Fig.\ \ref{fig:art_disp_travel}. Nevertheless, the formalism we derived describes the center of mass motion of the skyrmion qualitatively correctly, as illustrated by the comparison with the results of our simulations (both in a bilayer and in a single layer AFM) by simply re-scaling the value of the parameter $\Lambda$. Internal skyrmion modes are by construction not included in the Thiele approach. 

	\section{Conclusions} 
	\label{sec:conclusion}
	In this work, we show that small displacements between the two sublattice ferromagnetic constituents of an antiferromagnetic skyrmion induce a motion of the skyrmion. The concept of displacements reveals the mechanism how the effective mass of the AFM skyrmion is generated. We thereby answer the question how the inertia of an AFM skyrmion is rooted in the coupling between two sublattices. Furthermore, the small displacement of the two sublattice components generates a potential (or, tension) energy which is transformed to kinetic energy of the moving skyrmion. Spin polarized electric currents can cause such displacements between the two sublattice constituents of AFM skyrmions. For unidirectional sublattice currents, we find opposite skyrmion Hall effects for the two ferromagnetic constituents, which try to displace both in opposite directions. Their ensuing longitudinal velocities, induced by the current as well as by the displacement, add up to the resulting skyrmion velocity. Antiparallel sublattice currents, on the other hand, lead to skyrmion motion in perpendicular direction, solely caused by the displacement. In total, the motion finally corresponds to that of a classical driven, damped particle. Our analysis in terms of sublattice displacements explains that, whenever coupled FM skyrmions of opposite orientations move in different directions, such that their distance $\bm{\delta}$ increases, this adds a component perpendicular to $\bm{\delta}$ to the motion of the corresponding AFM skyrmion.\\

	\begin{acknowledgments}
			This work is funded by the Cluster of Excellence ``CUI: Advanced Imaging of Matter'' of the Deutsche Forschungsgemeinschaft (DFG) - EXC 2056 - project ID 390715994 (M.L. and M.T.). W.H. acknowledges support from the DFG SPP 2137 (skyrmionics) - project ID 403505707.
	\end{acknowledgments}
	
	\appendix
	\section{General derivation of displacement force and energy}
	\label{apendix1}
	The energy functional of the antiferromagnetic lattice consists of two ferromagnetic sublattices, $W_{\text{FM}}[\bm{a},\bm{a}]$ and $W_{\text{FM}}[\bm{b},\bm{b}]$, coupled antiferromagnetically by $W_{\text{c}}[\bm{a}, \bm{b}]$. Let us assume that the energy functionals $W[\bm{.},\bm{.}]$ can be written as bilinear forms of their arguments, e.g., $\bm{a}\cdot \bm{b}$, $\bm{a} \cdot \bm{\nabla}^2 \bm{b}$ or $\bm{a} \cdot (\bm{\nabla} \times \bm{b})$, including possible gradients. By symmetry, the coupling energy $W_{\text{c}}$ has to be invariant under sublattice exchange $\bm{a} \leftrightarrow \bm{b}$. Owing to translation invariance any displacement between sublattice magnetic structures will not alter the ferromagnetic parts of the energy. In the spirit of Thiele's approximation, we therefore ignore ferromagnetic terms when imposing sublattice displacements and focus on the coupling part of the energy functional. Applying a Taylor expansion to the fields $\bm{a}$ and $\bm{b}$, as described in the main text (note that $\bm{b}$ is displaced by $\bm{\delta}$ while, after $\bm{a} \leftrightarrow \bm{b}$ interchange, $\bm{a}$ is displaced by $-\bm{\delta}$) yields for the coupling energy
	\begin{equation}
		\label{eq:sm_coupling_energy_1}
		\begin{aligned}
			W_c[\bm{a}, \bm{b}] = &-W_c[\bm{b}, \bm{b}] + \sum_\mu \delta_\mu W_c[\partial_{\mu} \bm{b}, \bm{b}]\\ 
			&- \frac{1}{2} \sum_{\mu, \nu} \delta_\mu \delta_\nu W_c[\partial_{\mu} \partial_{\nu} \bm{b}, \bm{b}] \, ,
		\end{aligned}
	\end{equation}
	and
	\begin{equation}
		\label{eq:sm_coupling_energy_2}
		\begin{aligned}
			W_c[\bm{b}, \bm{a}] = &-W_c[\bm{a}, \bm{a}] - \sum_\mu \delta_\mu W_c[\partial_{\mu} \bm{a}, \bm{a}]\\ 
			&- \frac{1}{2} \sum_{\mu, \nu} \delta_\mu \delta_\nu W_c[\partial_{\mu} \partial_{\nu} \bm{a}, \bm{a}]\ .
		\end{aligned}
	\end{equation}
	Both second terms linear in $\bm{\delta}$ on the right hand sides must vanish due to sublattice exchange symmetry. The first term, $-W_c[\bm{b}, \bm{b}]$, is the coupling energy of a resting AFM skyrmion, i.e., of two exactly anti-parallel sublattice skyrmions. Due to the rotational invariance of the skyrmion, we conclude that the last term $W_c[\partial_{\mu} \partial_{\nu} \bm{b}, \bm{b}]$ vanishes for $\mu \neq \nu$ and is independent of whether $\mu=x$ or $\mu=y$. Thus, the total energy is $W = W_{\text{FM}}^A + W_{\text{FM}}^B - W_{\text{c}}[\bm{b}, \bm{b}] + \Delta W$, the energy of a resting AFM skyrmion plus some excitation energy $\Delta W = \frac{1}{2} |\bm{\delta}|^2 \Lambda$, where we define a constant $\Lambda := -W_c[\partial_{x} \partial_{x} \bm{b}, \bm{b}]$.
	
	In the following, we show that this quantity $\Lambda$ is the same force constant as obtained in the main text from analyzing the skyrmion dynamics. For a proof, we have to show that $\int \left(\partial_\mu \bm{a} \right) \cdot \bm{H}_{\text{eff}}^A \text{d}^2 \bm{r} = -\delta_\mu W_c[\partial_{\nu} \partial_{\nu} \bm{b}, \bm{b}]$, see the discussion of $\Lambda$ above and Eq.\ \eqref{eq:lambda_def}. The effective field is given by the derivative of the energy with respect to the corresponding sublattice field, $\bm{H}_{\text{eff}}^A = -\delta W / \delta \bm{a}$ or $\bm{H}_{\text{eff}}^A = -\delta W_{\text{FM}}[\bm{a}, \bm{a}] / \delta \bm{a} - \delta W_c[\bm{a}, \bm{b}] / \delta \bm{a}$. The ferromagnetic part of $\bm{H}_{\text{eff}}^A$, $-\delta W_{\text{FM}}[\bm{a}, \bm{a}] / \delta \bm{a}$, always points parallel to the magnetization field $\bm{a}$ for stationary magnetic structures \cite{LL}, such as the FM skyrmion on sublattice A. Therefore, $\int (\partial_{\mu} \bm{a} \cdot \delta W_{\text{FM}}[\bm{a}, \bm{a}] / \delta \bm{a}) \text{d}^2 \bm{r}=0$.
	
	For the coupling part to $\bm{H}_{\text{eff}}^A = - \delta W_c[\bm{a}, \bm{b}] / \delta \bm{a}$, we have $\int (\partial_{\mu} \bm{a} \cdot \delta W_c[\bm{a}, \bm{b}] / \delta \bm{a}) \text{d}^2 \bm{r}= W_c[\partial_{\mu} \bm{a}, \bm{b}]$ as can be deduced from $\frac{\delta}{\delta\bm{m}}\int (\bm{m} \cdot \delta W_c[\bm{a}, \bm{b}] / \delta \bm{a}) \text{d}^2 \bm{r}= \bm{H}_{\text{eff}}^A[\bm{b}]=\frac{\delta}{\delta\bm{m}}W_c[\bm{m}, \bm{b}]$, valid for any arbitrary magnetization field $\bm{m}$ on the sublattice A, i.e., in particular for $\bm{m}=\partial_{\mu} \bm{a}$. For $W_c[\partial_{\mu} \bm{a}, \bm{b}]$ we can again apply the Taylor expansion and get $W_c[\partial_{\mu} \bm{a}, \bm{b}] = -W_c[\partial_{\mu} \bm{b}, \bm{b}] + \sum_\nu \delta_\nu W_c[\partial_{\nu} \partial_{\mu} \bm{b}, \bm{b}] - \frac{1}{2} \sum_{\nu, \nu'} \delta_\nu \delta_{\nu'} W[\partial_{\nu} \partial_{\nu'} \partial_{\mu} \bm{b}, \bm{b}]$. Only the term linear in $\bm{\delta}$ survives, leading to $\int \left(\partial_{\mu} \bm{a} \right) \cdot \bm{H}_{\text{eff}}^{A}\ \text{d}^2 \bm{r} = - \delta_{\mu} W_c[\partial_{\nu} \partial_{\nu} \bm{b}, \bm{b}]$. Thus, the force constant $\Lambda$ for the skyrmion dynamics is, indeed, the same as the one derived previously from energy considerations.

	\section{Explicit integrals}
	\label{appendix2}
	Since isotropic skyrmions are characterized by their radial profile $\theta(\rho)$, the force constant
	\begin{equation}
		\label{eq:sm_Lambda}
		\begin{aligned}
			\Lambda &= \pi \int_{0}^{\infty} \text{d}\rho\ \bigg\{ A \rho  \theta ''(\rho )^2 - \frac{2 A \theta ''(\rho ) \sin (\theta (\rho )) \cos (\theta (\rho ))}{\rho }\\ 
			&- \frac{A \theta '(\rho ) \sin (2 \theta (\rho ))}{\rho^2} + A \rho  \theta '(\rho )^4 + \frac{2 A \theta '(\rho )^2}{\rho }\\ 
			&- \frac{A \theta '(\rho )^2 \cos (2 \theta (\rho ))}{\rho} + 2 A \theta '(\rho ) \theta ''(\rho ) + \frac{A \sin ^2(\theta (\rho ))}{\rho^3}\\ 
			&+ D \theta ''(\rho ) \sin ^2(\theta (\rho )) + D \rho  \theta '(\rho )^3 + \frac{2 D \theta '(\rho ) \sin ^2(\theta (\rho ))}{\rho }\\ 
			&+ \frac{1}{2} D \theta '(\rho )^2 \sin (2 \theta (\rho )) + \frac{1}{2} \lambda  \rho  \theta '(\rho )^2 + \frac{\lambda  \sin ^2(\theta (\rho ))}{2 \rho }\\ 
			&- K \rho  \theta ''(\rho ) \sin (\theta (\rho )) \cos (\theta (\rho )) - \frac{1}{2} K \rho  \theta '(\rho )^2\\ 
			&- \frac{1}{2} K \theta '(\rho ) \sin (2 \theta (\rho )) - \frac{1}{2} K \rho  \theta '(\rho )^2 \cos (2 \theta (\rho )) \bigg\}\ ,
		\end{aligned}
	\end{equation}
	and the magnitude of the dissipation tensor
	\begin{equation}
		\label{eq:sm_D}
		\gamma \mathcal{D} = \int_{0}^{\infty} \text{d}\rho \left\{ \pi  \rho  \theta '(\rho )^2+\frac{\pi  \sin ^2(\theta (\rho ))}{\rho } \right\} 
	\end{equation}
	can be provided as integrals over $\rho$. Eq.\ \eqref{eq:sm_D} agrees with Eq.\ (14) in Ref.\ \cite{IwaUniversal}. The remaining integral over $\rho$ can be carried out for explicit expressions of the skyrmion profile $\theta(\rho)$, or numerically. By considering the form $\theta(\rho)$ of Eq.\ \eqref{eq:fit_theta}, and fitting the skyrmion used in the simulations, we obtain a radius $R/d=13.3$ and a width $w/d=5.6$ which finally yields $\mathcal{D} \approx 18.5$ for the dissipation strength.

\end{document}